\documentclass[review]{elsarticle}

\usepackage[margin=1in]{geometry}
\usepackage{float}
\usepackage{lineno,hyperref}
\usepackage{color}
\usepackage{amssymb}
\usepackage{graphics}
\usepackage{graphicx}
\usepackage{pbox}
\usepackage{mhchem}
\usepackage{xcolor} 
\usepackage[ampersand]{easylist}
\usepackage{lineno}
\usepackage{booktabs}
\usepackage{multirow}
\usepackage{subcaption}
\usepackage{amsmath}
\usepackage{paralist}
\usepackage{soul}
\usepackage{siunitx}
\usepackage{amsmath}
\usepackage{url}
\usepackage{makecell}
\usepackage{hyperref}
\usepackage{enumitem}

\newcommand{\argmin}{\mathop{\mathrm{argmin}}\limits}
\newcommand{\solve}{\mathop{\text{Solve}}\limits}
\newcommand{\cm}{\text{cm}} 
\newcommand{\g}{\text{g}}

\newcommand{\eff}{\text{eff}}

\newcommand{\PE}{\text{PE}}
\newcommand{\CS}{\text{CS}}
\newcommand{\PP}{\text{PP}}

\hypersetup{  
    colorlinks=true,
    linkcolor=blue,
    filecolor=magenta,      
    urlcolor=cyan,
    pdftitle={Sharelatex Example},
    bookmarks=true,
    pdfpagemode=FullScreen,
    citecolor=blue
    }

\modulolinenumbers[1]

\makeatletter
\def\ps@pprintTitle{%
 \let\@oddhead\@empty
 \let\@evenhead\@empty
 \def\@oddfoot{\centerline{\thepage}}%
 \let\@evenfoot\@oddfoot}
\makeatother

\begin{document}

\begin{frontmatter}

\title{Atomic number estimation of dual energy cargo radiographs using a semiempirical transparency model}

\author[MITaddress]{Peter Lalor\corref{corauthor}}
\cortext[corauthor]{Corresponding author 
\\\hspace*{13pt} Email address: plalor@mit.edu
\\\hspace*{13pt} Telephone: (925) 453-1876 
\\\hspace*{13pt} 138 Cherry St, Cambridge, MA 02139}

\author[MITaddress]{Areg Danagoulian}

\address[MITaddress]{Department of Nuclear Science and Engineering, Massachusetts Institute of Technology, Cambridge, MA 02139, USA}

\begin{abstract}
Dual energy cargo inspection systems are sensitive to both the area density and the atomic number of an imaged container due to the $Z$ dependence of photon attenuation. The ability to identify cargo contents by their atomic number enables improved detection capabilities of illicit materials. This work introduces a novel method for atomic number reconstruction by minimizing the chi-squared error between measured transparency values and a semiempirical transparency model. This method is tested using two Geant4 Monte Carlo simulated radiographic phantoms, demonstrating the ability to obtain accurate material predictions on noisy input images, even in the presence of shielding. Furthermore, we provide a simple procedure for porting this method to a commercial system, requiring an approximate model of the scanner's beam spectra and detector response, along with only three calibration measurements.
\end{abstract}
\end{frontmatter}
\begin{sloppypar}


\section{Introduction}
\label{Introduction}

Every year, over 32.7 million cargo containers enter through U.S. ports~\cite{CBP2021}. A potential security concern is that a terrorist could smuggle nuclear material through U.S. ports and subsequently assemble a nuclear weapon~\cite{Cochran2008}. A nuclear detonation at a U.S. port could result in excess of \$1 trillion in net economic costs, including infrastructure damage and trade disruption~\cite{Meade2006, Rosoff2007}. To combat these threats, the U.S. Congress passed the SAFE Port Act in 2006, which mandated 100 percent screening of U.S. bound cargo and 100 percent scanning of high-risk containers~\cite{PLAW109-347}. The U.S. shortly afterwards passed the Implementing Recommendations of the 9/11 Comission Act of 2007, invoking a 2012 deadline for a full-scale implementation requiring that all containers be scanned prior to entering the U.S.~\cite{PLAW110-53}. This deadline continues to be extended due to lack of technological solutions.

The U.S. scans all high risk containers (identified as approximately 5 percent of seaborne containers~\cite{CBO2016}) using non-intrusive inspection (NII) technology~\cite{NII}. These radiography systems measure the attenuation of X-rays and/or gamma rays which are directed through the container to produce an attenuation image of the scanned cargo. Some radiography systems deploy dual energy photon beams, enabling classification of objects according to their $Z$, since the attenuation of photons depends on the atomic number of the intervening material. This technology improves the capabilities of these systems to identify nuclear threats or high-$Z$ shielding.

This work presents a novel method for predicting the area density and atomic number of dual energy radiographic images. In Section~\ref{Background}, we provide background surrounding dual energy radiography and discuss current techniques. In Section~\ref{Methodology}, we describe our approach for $Z$ estimation. In Section~\ref{Porting}, we outline the steps necessary to port this algorithm to commercial scanners. In Section~\ref{Analysis}, we describe the results of our method using two Monte Carlo simulated radiographic images. In Section~\ref{Conclusion}, we summarize the advantages of our method and discuss future work.

\section{Background}
\label{Background}

\subsection{Dual energy radiography overview}

When a radiography system scans a material of area density $\lambda$ and atomic number $Z$, it measures the transparency of the photon beam, defined as the detected charge in a scintillator-based sensor in the presence of the material normalized by the open beam measurement. Using the Beer-Lambert law, the photon transparency $T(\lambda, Z)$ can be expressed as follows:

\begin{equation}
T(\lambda, Z) = \frac{\int_0^{\infty} D(E) \phi(E) e^{-\mu (E, Z) \lambda} dE}{\int_0^{\infty}D(E) \phi(E) dE}.
\label{transparency}
\end{equation}

In  Eq.~\ref{transparency}, $\mu(E, Z)$ is the mass attenuation coefficient, $\phi(E)$ is the differential photon beam spectrum, and $D(E)$ is the detector response function. For this work, we calculate the photon beam spectrum and detector response function from the output of Geant4 simulations, as described in \ref{simulation}~\cite{Geant4, grasshopper}. The radiographic transparency is dependent on the atomic number of the imaged material through the mass attenuation coefficient $\mu(E, Z)$. Thus, by making multiple transparency measurements of the same object using different photon energy spectra, properties of the material $Z$ can be inferred.

\subsection{Limitations of existing methods}
\label{limitations}

Existing dual energy atomic number reconstruction methods can be categorized into two approaches: analytic methods and empirical methods. Analytic methods obtain material predictions by comparing radiographic measurements to model predictions~\cite{Novikov1999, Zhang2005, Ogorodnikov2002, Li2016}. However, this approach is generally unsuitable for practical applications, since the accuracy of Eq.~\ref{transparency} deteriorates significantly under non-ideal circumstances. For instance, Eq.~\ref{transparency} assumes only noninteracting photons are detected, and thus ignores the effects of scattered radiation, which can be up to one percent of the detected X-ray beam~\cite{Chen2007}. Furthermore, Eq.~\ref{transparency} requires a high precision model for the photon beam spectra and detector response function, which may not be known exactly for real applications.

Other authors instead perform material discrimination using an empirical calibration step~\cite{Budner2006, Perticone2010, Lee2012, Lee2018}. Empirical approaches are more accurate than analytic methods because the transparency model is constructed using calibration scans, reducing inherent model bias. However, the calibration step can be time consuming, posing a challenge if a scanning system needs to be recalibrated. Furthermore, these methods can only identify materials according to a small number of material classes, introducing binning bias and limiting atomic number selectivity. Modified approaches attempt to make finer atomic number predictions by interpolating between different material classes, although this introduces a source of error when the $Z$ of an object differs significantly from the calibration materials.

\subsection{Calibrating the semiempirical transparency model}

Past work found that the limitations of Eq.~\ref{transparency} can be accurately corrected for by replacing the mass attenuation coefficient with a semiempirical mass attenuation coefficient $\tilde \mu(E, Z)$~\cite{Lalor2023}:

\begin{equation}
\tilde \mu(E, Z) = a \mu_\PE(E, Z) + b \mu_\CS(E, Z) + c \mu_\PP(E, Z) \\
\label{semiempirical_mass_atten}
\end{equation}

where $\mu_\PE(E, Z)$, $\mu_\CS(E, Z)$, and $\mu_\PP(E, Z)$ are the mass attenuation coefficients from the photoelectric effect (PE), Compton scattering (CS), and pair production (PP), calculated from NIST cross section tables~\cite{NIST}. In Eq.~\ref{semiempirical_mass_atten},  $a$, $b$, and $c$, are determined through a calibration step, as described in \ref{calculating_abc}. We emphasize that the calibration step only requires measurements of a single thickness of three different materials, which is far simpler than the calibration required by typical commercial systems. After substituting Eq.~\ref{semiempirical_mass_atten} into Eq.~\ref{transparency}, we refer to this improved transparency model as the semiempirical transparency model.

\subsection{Calculating ground truth $Z_\eff$}
\label{calc_ground_truth}

Even if the precise material composition of a detected object is known exactly, determining the ground truth effective atomic number $Z_\eff$ is a somewhat unclear task if the object is not homogeneous. Past authors have proposed definitions of $Z_\eff$ involving a weighted average of atomic numbers~\cite{Naydenov2013, Langeveld2017_effective_Z}. However, as described in prior work, such definitions lead to inconsistencies~\cite{Lalor2024}. Instead, the most precise definition for $Z_\eff$ is the atomic number of a homogeneous material that would produce identical high and low energy transparency measurements as the heterogeneous material. Thus, to solve for the ground truth $Z_\eff$ of a known array of materials, this work uses the semiempirical transparency model to find the $\{\lambda_\eff, Z_\eff\}$ which produces the same transparency predictions as the heterogeneous material. This procedure is discussed in more detail in \ref{calc_ground_truth_appendix}. While this approach is only approximate, we expect it to yield an accurate result for algorithmic benchmark purposes.

\section{Methodology}
\label{Methodology}

\subsection{Image segmentation}

Pixel-by-pixel atomic number reconstruction is impractical due to the poor signal-to-noise ratio of individual pixels. Instead, the image is first segmented into different regions and subsequently the material properties of each cluster are predicted. Past studies describe various segmentation methods for cargo applications, including a modified leader algorithm and a hybrid k-means region growing technique~\cite{Ogorodnikov2002, Fu2010}. This study uses Felzenszwalb's image segmentation algorithm, chosen due to its speed, flexibility, and minimal hyperparameter tuning~\cite{Felzenszwalb2004}. The algorithm to be described in the next section was performed independently on every segment. For practical applications, a specialized image segmentation routine should be used, as the accuracy of this work is highly dependent on the ability to obtain a precise segmentation.

\subsection{Atomic number reconstruction}
\label{Z_recon}

We define a new coordinate, $\alpha$, using a log transform $\alpha = - \log T$. Furthermore, we use the subscripts subscripts $\{H,~L\}$ to distinguish between the \{high, low\} energy beam measurements. Using this notation, we define $\{\alpha_{H,k},~\alpha_{L,k}\}$ as the measured log-transparencies for pixel $k$ in a pixel cluster $C$, with corresponding uncertainties $\{\sigma_{\alpha_{H,k}},~\sigma_{\alpha_{L,k}}\}$. Furthermore, we define $(\lambda_k,Z)$ as the area density and atomic number estimates of pixel $k$, imposing that every pixel in the segment is assigned the same atomic number. Using the semiempirical log-transparency model, given explicitly by Eq.~\ref{alpha}, we write the estimated log-transparencies of pixel $k$ as $\{\tilde \alpha_H (\lambda_k, Z),~\tilde \alpha_L (\lambda_k, Z)\}$. A chi-squared objective function can then be defined as follows:

\begin{equation}
\chi^2( \boldsymbol{\lambda}, Z) = \sum_{k \in C} \left[ \frac{\left(\tilde \alpha_H(\lambda_k, Z) - \alpha_{H,k} \right)^2}{\sigma_{\alpha_{H,k}}^2} + \frac{\left(\tilde \alpha_L (\lambda_k, Z) - \alpha_{L,k} \right)^2}{\sigma_{\alpha_{L,k}}^2} \right].
\label{chi2}
\end{equation}

The area density and atomic number predictions for each pixel in $C$ are thus chosen as the minimizer of Eq.~\ref{chi2}. \ref{optimization} describes in detail how to efficiency minimize Eq.~\ref{chi2} using Newton's method.

\section{Porting this approach to commercial systems}
\label{Porting}

The methodology described in this work can easily be adapted to commercial systems. The steps can be summarized as following:

\begin{enumerate}[label=(\arabic*)]
\item Calculate an approximate model of the system's beam energy spectra $\phi_H(E)$ and $\phi_L(E)$, and detector response function $D(E)$, using either Monte Carlo simulations or experimental measurements.
\item Perform at least three experimental calibration measurements, recording the true $\lambda$ and $Z$ of the calibration objects along with the measured log-transparencies $\alpha_H$ and $\alpha_L$. 
\item Use Eq.~\ref{calc_abc} to calculate the calibration parameter values which best reproduce the calibration measurements through a least-squares routine. The forward model is now defined by Eq.~\ref{alpha}.
\item To compute the pixel-by-pixel $Z_\eff$ estimate of a radiographic image, first perform an image segmentation step to group similar pixels. Then, for each pixel segment, minimize Eq.~\ref{chi2} using the method described in \ref{optimization}.
\end{enumerate}

Steps (1), (2), and (3) only need to be performed once. Steps (2) and (3) should be repeated if the system needs to be recalibrated, but do not need to be repeated between different container scans. A significant advantage of this approach is the simplicity of the calibration step, requiring only three calibration scans, as described in \ref{calculating_abc}. Step (4) is then performed independently on each radiographic image. We emphasize that implementing this methodology does not require a detailed simulated model of the scanning system.

\section{Analysis}
\label{Analysis}

\subsection{Cargo phantom}
\label{Cargo_phantom}

In order to test the effectiveness of the atomic number reconstruction routine described in Section~\ref{Methodology}, this work ran Geant4 Monte Carlo simulations of a radiographic cargo phantom~\cite{Geant4,grasshopper}. Seven boxes composed of graphite ($Z=6$), aluminum ($Z=13$), iron ($Z=26$), silver ($Z=47$), gadolinium ($Z=64$), lead ($Z=82$), and uranium ($Z=92$) are placed inside a steel container. Below the boxes are cylinders of water (H$_2$O), silver chloride (AgCl), and uranium oxide (UO$_2$). Next to the cylinders is a plutonium pit ($Z=94$) surrounded by polyethylene shielding (with the reduced formula of CH$_2$). The simulation geometry is described in more detail in \ref{simulation_geom}. 

Fig.~\ref{Z_cargo_true} shows the ground truth $\{\lambda, Z\}$ map of the cargo phantom. When displaying $\{\lambda, Z\}$ maps of phantom containers, the image opacity shows the material area density and the colorbar identifies the material atomic number. Determining the ground truth atomic number sometimes yields a solution degeneracy in which two different materials produce the same transparency values~\cite{Lalor2024}, as discussed in more detail in \ref{calc_ground_truth_appendix}. In the case of non-unique solutions, we record both atomic number solutions, but only display the lower-$Z$ solution for plotting purposes, and maintain this convention throughout the paper.

\subsection{Noise model}
\label{Noise_model}

The presence of noise in a radiographic image presents a significant challenge when computing the $Z$ of different pixels due to the poor conditioning of the inversion. In real systems, the pixel noise stems from competing sources of statistical and systematic measurement uncertainty, so a simple noise model assuming Poisson statistics is insufficient. Instead, we turn to previous results from a data-driven uncertainty study by Henderson, finding that the measured uncertainty is modeled well by a low-order polynomial in $\alpha$~\cite{Henderson2019}. Motivated by this result, we resample simulated pixel values from a Gaussian with a width of $0.1\alpha$, equivalent to adding $10\%$ Gaussian noise to the raw simulation output. While this noise model is far simpler than real cargo scanning systems, it serves as a conservative, worst case scenario estimation for algorithmic benchmark purposes. Fig.~\ref{simulation_H_cargo_noisy} shows the Geant4 simulated high energy cargo phantom using the noise model described in this section.

\subsection{Atomic number reconstruction accuracy}
\label{Z_accuracy}

\begin{figure}
    \centering
    \begin{subfigure}[t]{0.49\textwidth}
        \centering
        \includegraphics[width=\textwidth]{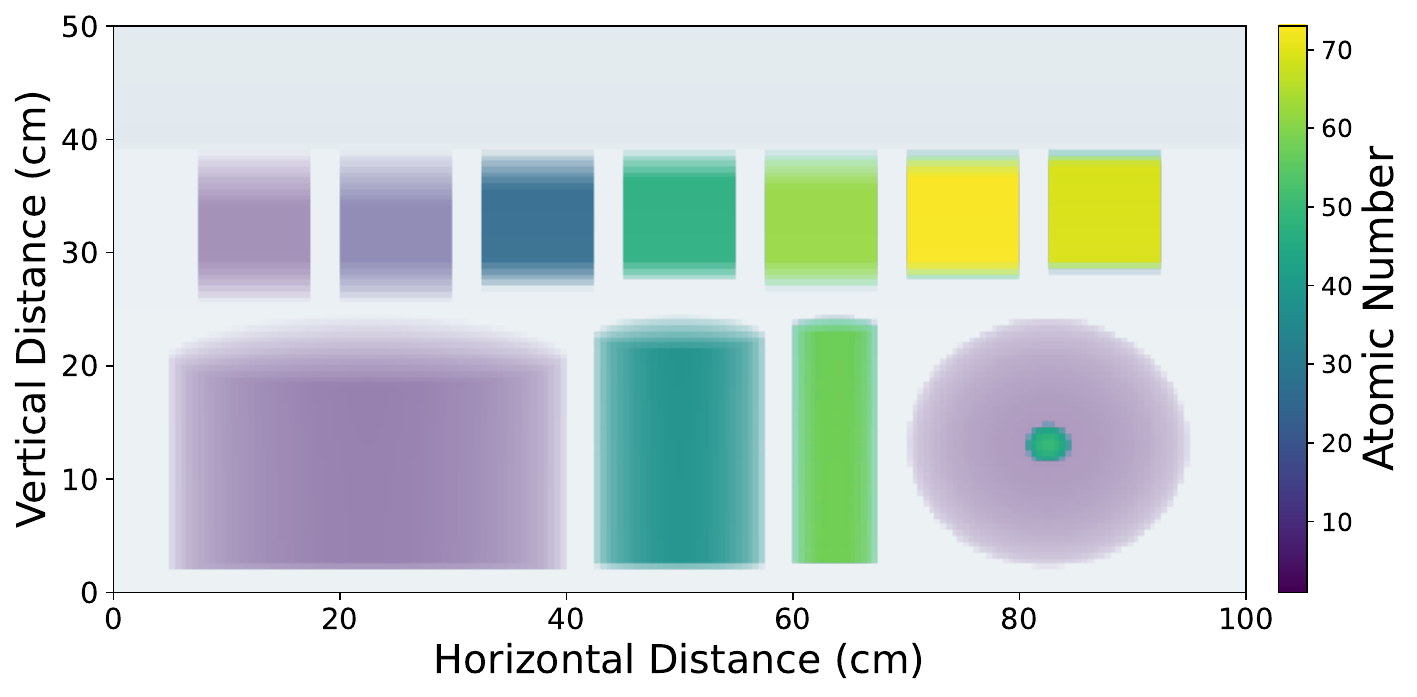}
        \caption{Ground truth $\{\lambda, Z\}$ map of the cargo phantom, calculated using Eq.~\ref{ground_truth}.}
        \label{Z_cargo_true}
    \end{subfigure}
   \hfill
    \begin{subfigure}[t]{0.49\textwidth}
        \centering
        \includegraphics[width=\textwidth]{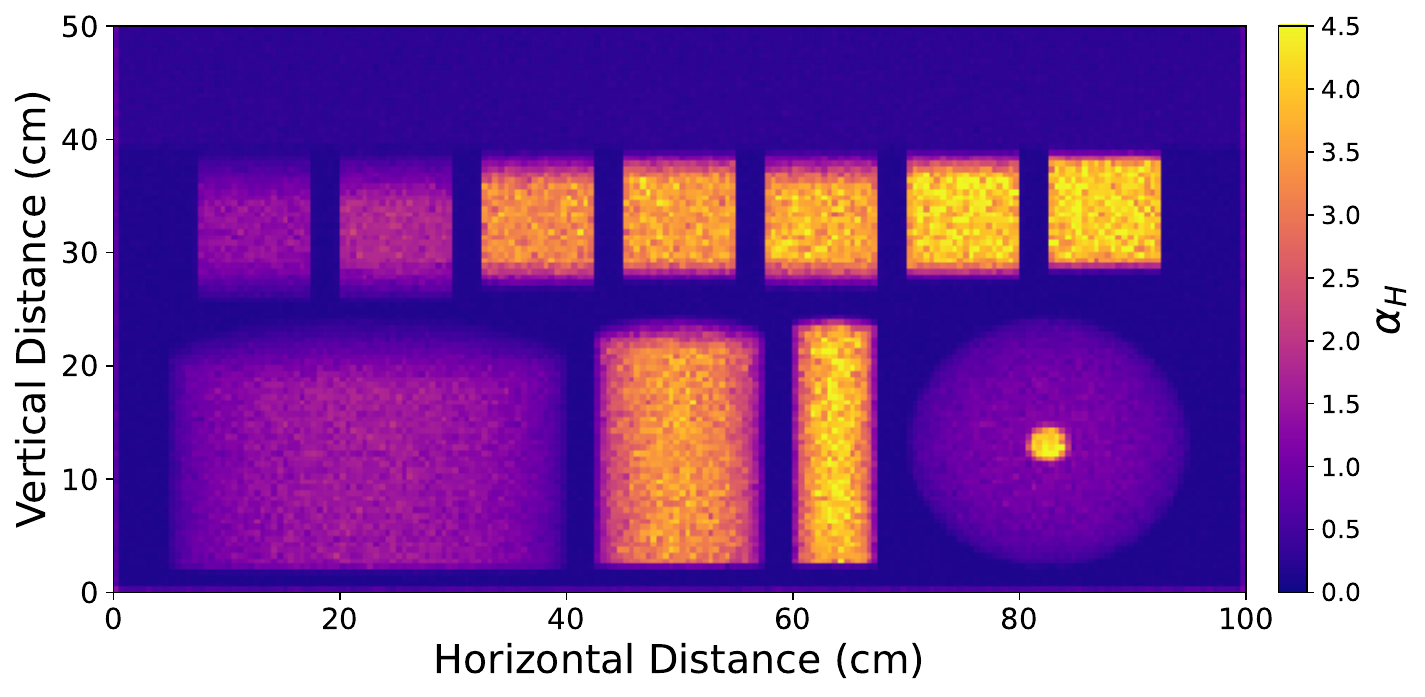}
        \caption{Geant4 simulated high energy cargo phantom using the noise model described in Section~\ref{Noise_model}.}
        \label{simulation_H_cargo_noisy}
    \end{subfigure}
    \vskip\baselineskip
    \begin{subfigure}[t]{0.49\textwidth}
        \centering
        \includegraphics[width=\textwidth]{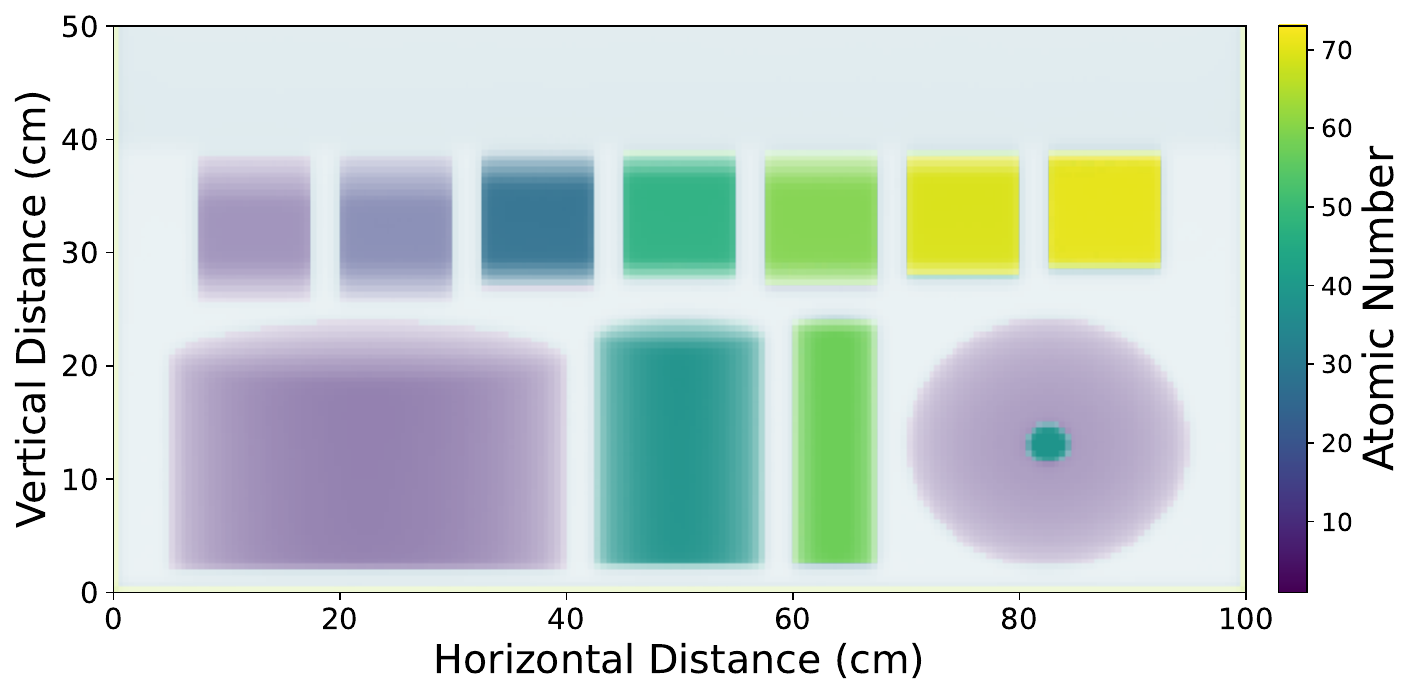}
        \caption{Reconstructed $\{\lambda, Z\}$ map of the noisy simulated cargo phantom, median across 1000 runs}
        \label{Z_cargo_med}
    \end{subfigure}
    \hfill
    \begin{subfigure}[t]{0.49\textwidth}
        \centering
        \includegraphics[width=\textwidth]{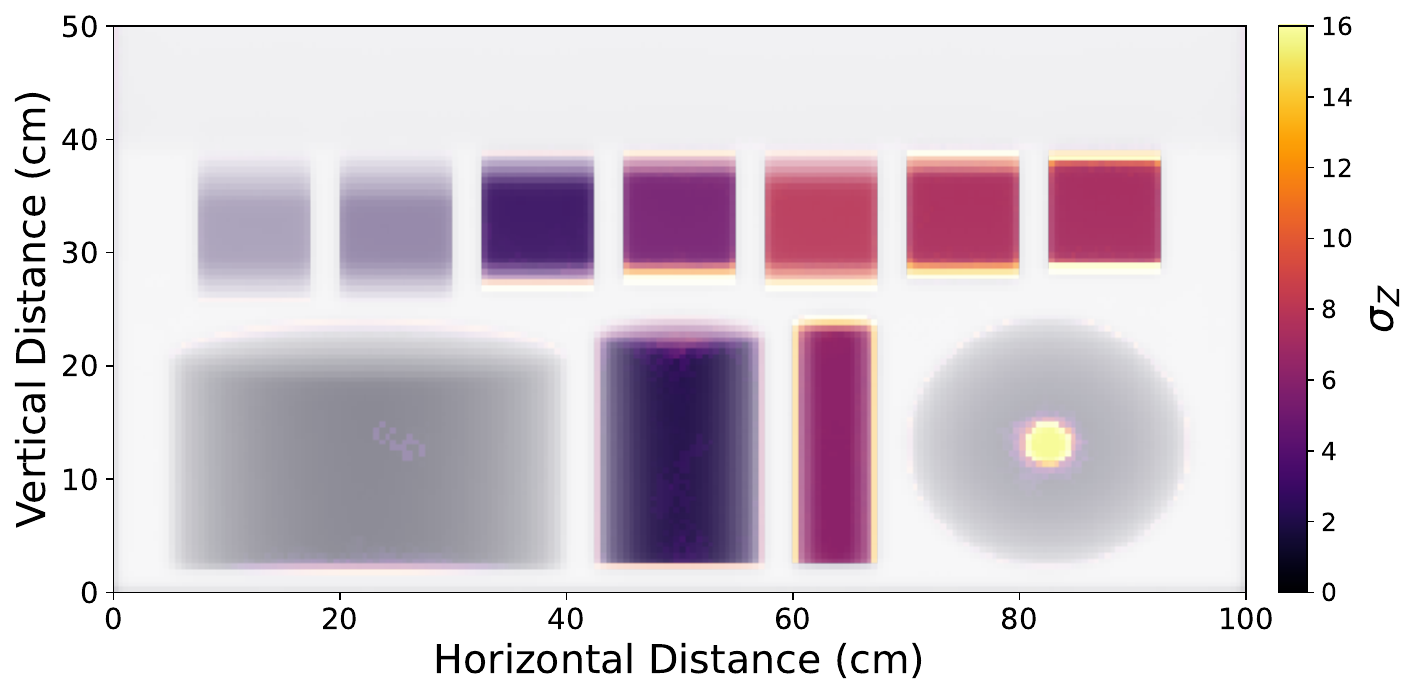}
        \caption{Atomic number reconstruction uncertainty of the noisy simulated cargo phantom, computed as the pixel-wise sample standard deviation of reconstructed $Z$ output across 1000 runs.}
        \label{Z_cargo_unc}
    \end{subfigure}
    \caption{Atomic number reconstruction results using the simulated radiographic cargo phantom.}
    \label{phantom}
\end{figure}

Using the simulated radiographic phantom as a template, 1000 noisy images were generated by resampling from the noise model described in Section~\ref{Noise_model}. Since this noise model reflects a conservative estimate of competing sources of statistical and systematic uncertainties, this choice fairly represents repeating a container scan 1000 times in order to benchmark algorithmic performance capabilities. The method presented in Section~\ref{Methodology} was then run independently on each of the noisy images to obtain pixel-wise area density and atomic number estimates. If multiple $Z$ solutions are found, we record both atomic number solutions (for further discussion, see~\ref{optimization}). The median and standard deviation of the 1000 runs is shown in Figures~\ref{Z_cargo_med} and \ref{Z_cargo_unc}, respectively.

Overall, the reconstructed atomic number image shows strong agreement with ground truth. We quantify these results in Table \ref{table_cargo_phantom}, where for every object of interest in the cargo phantom, we calculate the ground truth $Z_\eff$, reconstructed $Z$, and reconstructed $\sigma_Z$, defined as the average over all pixels in the object. In all cases, the reconstructed atomic number is within the uncertainty estimate of the ground truth $Z_\eff$. We notice that the atomic uncertainty is largest near horizontal material boundaries, primarily due to geometric effects near object peripheries in which photons pass through a much smaller fraction of the 3D object. We emphasize that $\sigma_Z$ is highly dependent on the level of noise of the input image, and applying this method to radiographs with better statistical resolution would yield lower values for $\sigma_Z$.

\begin{table}
\begin{centering}
\begin{tabular}{l c c}
\toprule
 & \makecell{Ground Truth $Z_\eff$ \\ (average per object)} & \makecell{Reconstructed $Z$ \\ (average per object)}  \\
\midrule
 Polyethylene shield & $7.6$ & $8.9 \pm 1.1$ \\
 Graphite box & $8.4$ & $10.3 \pm 2.1$ \\
 Water cylinder & $8.5$ & $9.1 \pm 0.8$ \\
 Aluminum box & $14.5$ & $16.2 \pm 2.2$ \\
 Iron box & $26.0$ & $27.3 \pm 3.2$ \\
 Silver chloride cylinder & $37.4$ & $37.9 \pm 2.3$ \\
 Plutonium pit + polyethylene shield & $46.0$ & $38.3 \pm 15.8$ \\
 Silver box & $45.9$ & $46.4 \pm 5.4$ \\
 Uranium oxide cylinder & $56.4$ & $57.2 \pm 6.3$ \\
 Gadolinium box & $61.1$ & $59.3 \pm 8.0$ \\
 Lead box & \{$72.1$, $92.7$\} & \{$69.0 \pm 7.5$, $86.2 \pm 8.6$\} \\
  Uranium box & \{$68.8$, $95.9$\} & \{$70.2 \pm 7.3$, $88.0 \pm 7.3$\} \\
\bottomrule
\end{tabular}
\caption{Reconstructing the $Z$ of different regions within the cargo phantom. For each region, the average value of $Z$ and the average value of $\sigma_Z$ are calculated. Note that the objects are all placed inside a steel container. Cells with two entries indicate the presence of a $Z$ degeneracy in which multiple solutions were found.}
\label{table_cargo_phantom}
\end{centering}
\end{table}

\subsection{Identification of shielded objects}
\label{identification_shielded}

A security vulnerability of dual energy radiographic systems arises from a smuggler's ability to effectively conceal high-$Z$ materials from detection by embedding them within low-$Z$ shielding. To combat this limitation, we observe that the closed form nature of the transparency model used in this work offers a unique way to identify shielded objects. If an object with area density $\lambda_\text{object}$ and atomic number $Z_\text{object}$ is obscured by shielding with area density $\lambda_\text{shield}$ and atomic number $Z_\text{shield}$, we can separate the semiempirical mass attenuation coefficient into two components:

\begin{equation}
\tilde \mu(E, Z) \lambda = \tilde \mu(E, Z_\text{object}) \lambda_\text{object} + \tilde \mu(E, Z_\text{shield}) \lambda_\text{shield}
\label{alpha_shielded}
\end{equation}

During a first pass, we approximate $(\lambda_\text{shield}, Z_\text{shield})$ using the methods described in Section~\ref{Methodology} by only considering the regions surrounding the object. We can then define a new forward model, replacing the mass attenuation coefficient with Eq.~\ref{alpha_shielded}, and then rerun the algorithm on the region of interest. This enables us to solve directly for $\{\lambda_\text{object}, Z_\text{object}\}$ during a second pass, exposing objects which are hidden behind the shielding and inferring their intrinsic values of $\lambda$ and $Z$. We note that in order for this method to be effective, one must be able to geometrically isolate the shielding during the first pass, which may not always be possible for heavily loaded or cluttered containers. For such scenarios, more advanced segmentation algorithms may be necessary, requiring future research in this area.

\subsection{Simulated shielded phantom}

\begin{figure}
    \centering
    \begin{subfigure}[t]{0.49\textwidth}
        \centering
        \includegraphics[width=\textwidth]{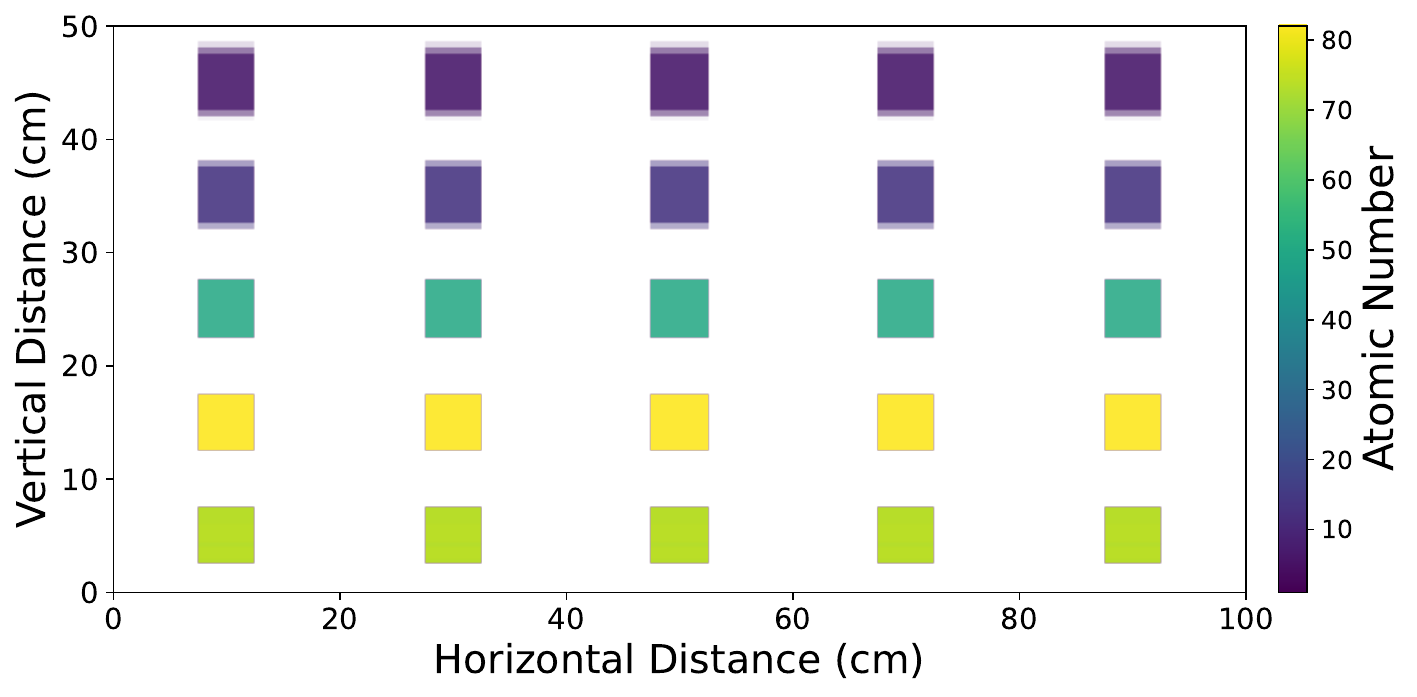}
        \caption{Ground truth $\{\lambda, Z\}$ map of the objects behind the steel shielding.}
        \label{Z_shielded_peeled_true}
    \end{subfigure}
    \hfill
    \begin{subfigure}[t]{0.49\textwidth}
        \centering
        \includegraphics[width=\textwidth]{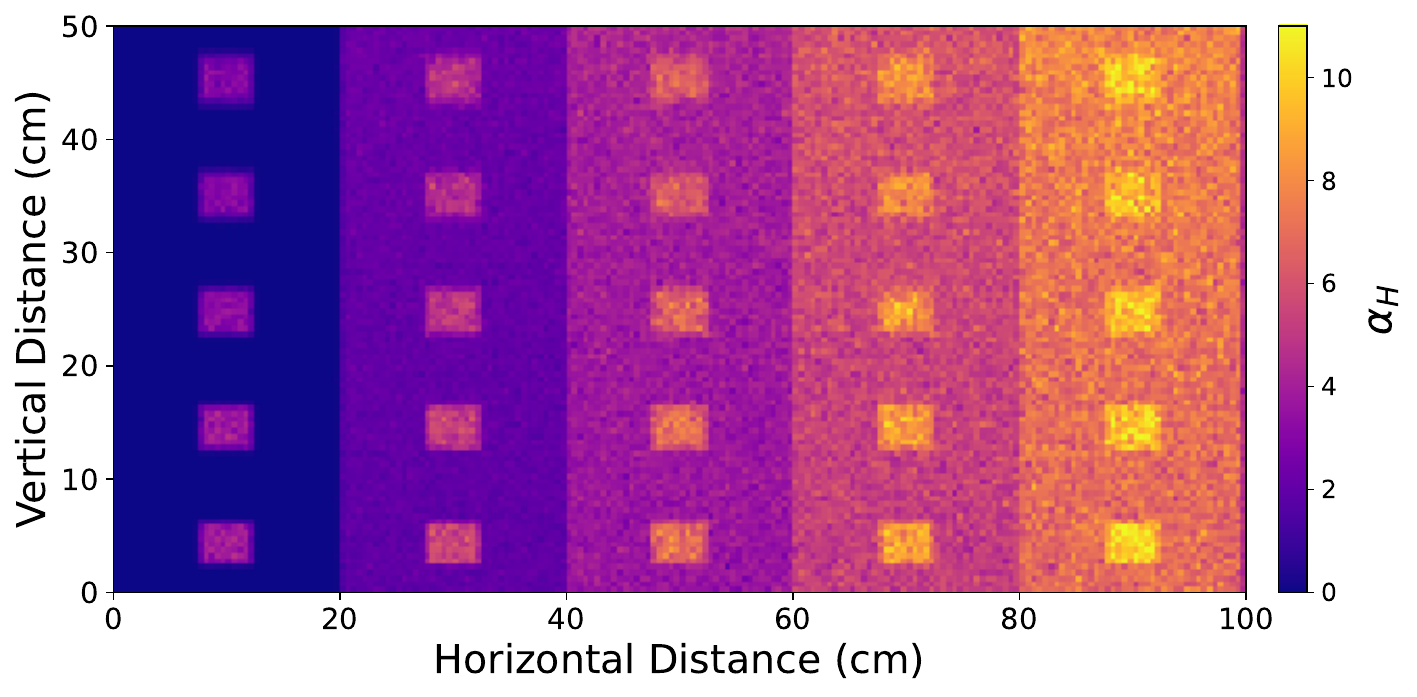}
        \caption{Geant4 simulated high energy shielded phantom using the noise model described in Section~\ref{Noise_model}.}
        \label{simulation_H_shielded_noisy}
    \end{subfigure}
    \vskip\baselineskip
    \begin{subfigure}[t]{0.49\textwidth}
        \centering
        \includegraphics[width=\textwidth]{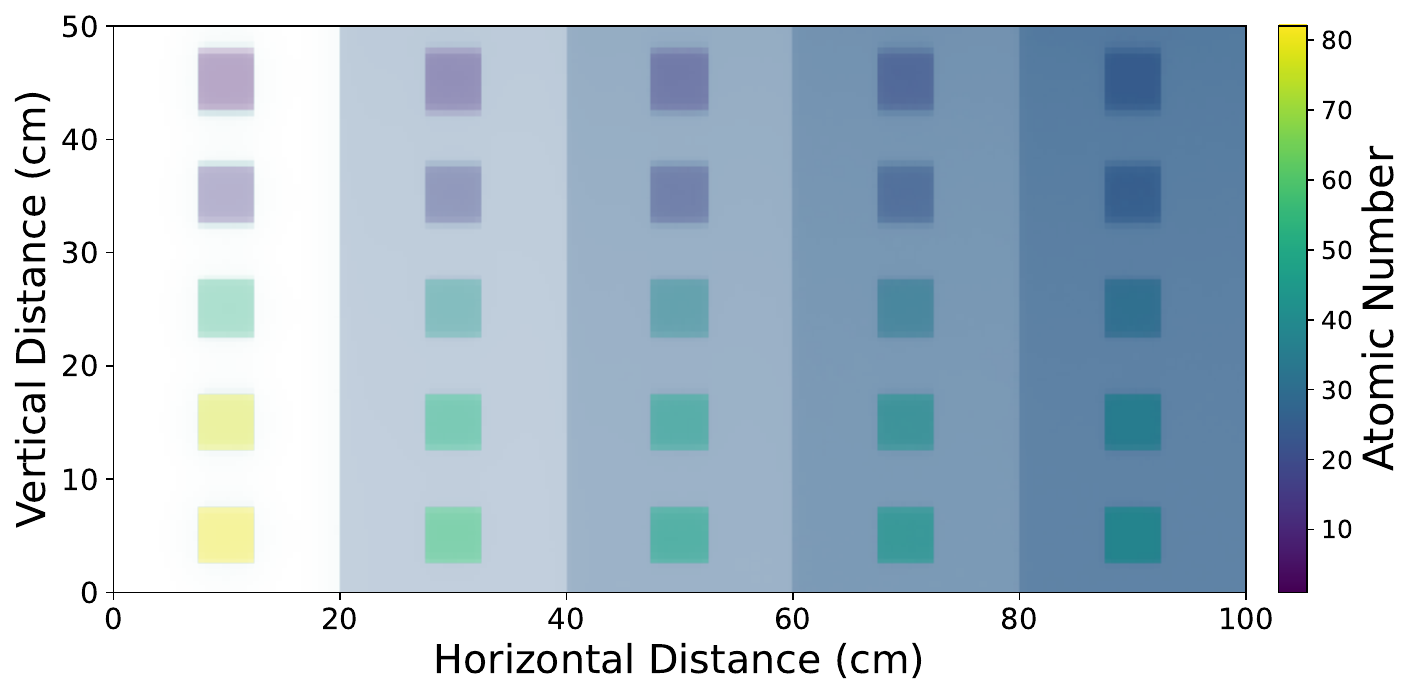}
        \caption{Reconstructed $\{\lambda, Z\}$ map of the noisy simulated shielded phantom without stripping the steel, median across 1000 runs.}
        \label{Z_shielded_med}
    \end{subfigure}
    \hfill
    \begin{subfigure}[t]{0.49\textwidth}
        \centering
        \includegraphics[width=\textwidth]{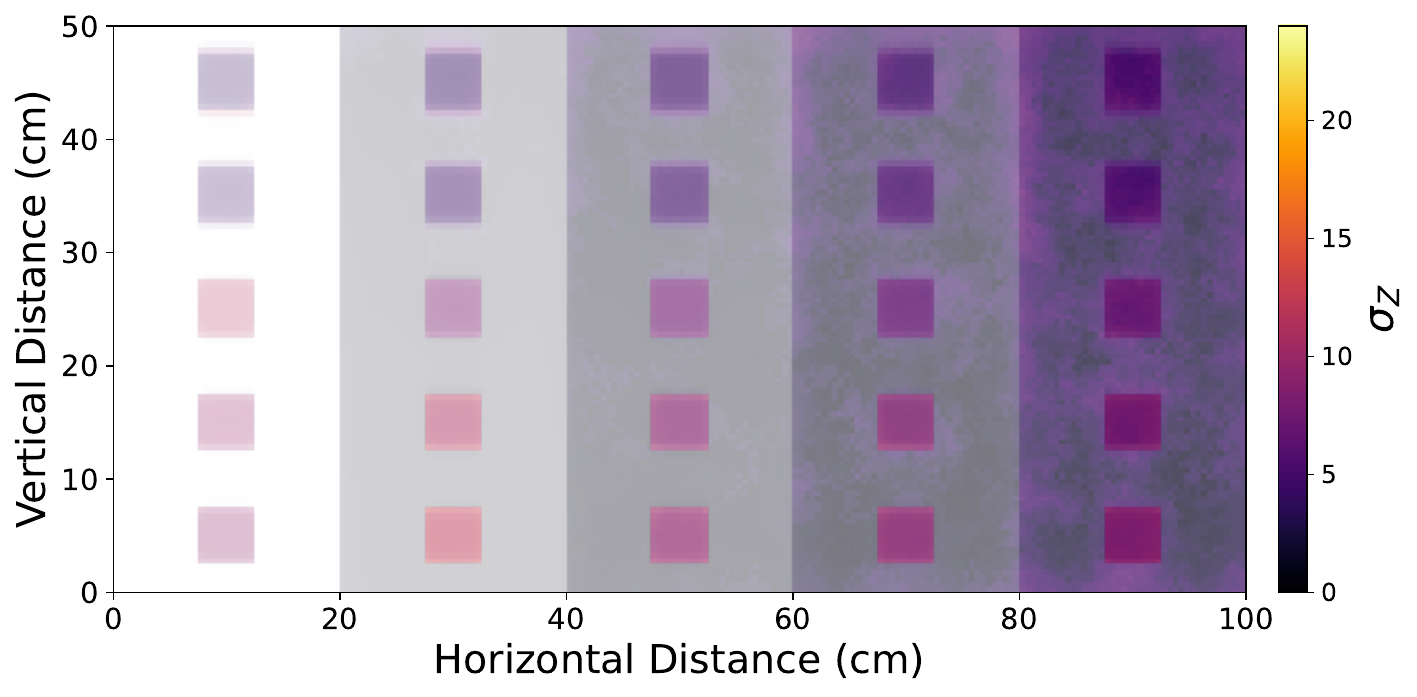}
        \caption{Atomic number reconstruction uncertainty of the noisy simulated shielded phantom without stripping the steel, computed as the pixel-wise sample standard deviation of reconstructed $Z$ output across 1000 runs.}
        \label{Z_shielded_unc}
    \end{subfigure}
    \vskip\baselineskip
    \begin{subfigure}[t]{0.49\textwidth}
        \centering
        \includegraphics[width=\textwidth]{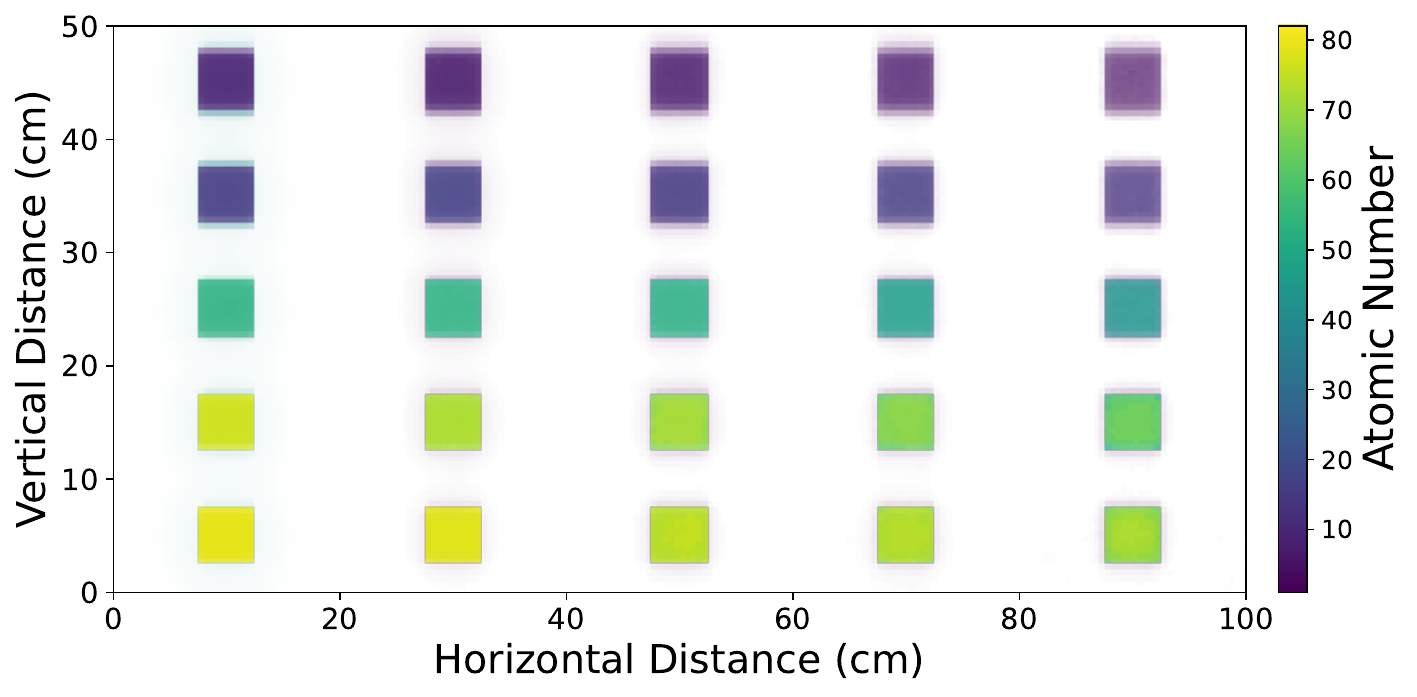}
        \caption{Reconstructed $\{\lambda, Z\}$ map of the noisy simulated shielded phantom after stripping the steel, median across 1000 runs.}
        \label{Z_shielded_peeled_med}
    \end{subfigure}
    \hfill
    \begin{subfigure}[t]{0.49\textwidth}
        \centering
        \includegraphics[width=\textwidth]{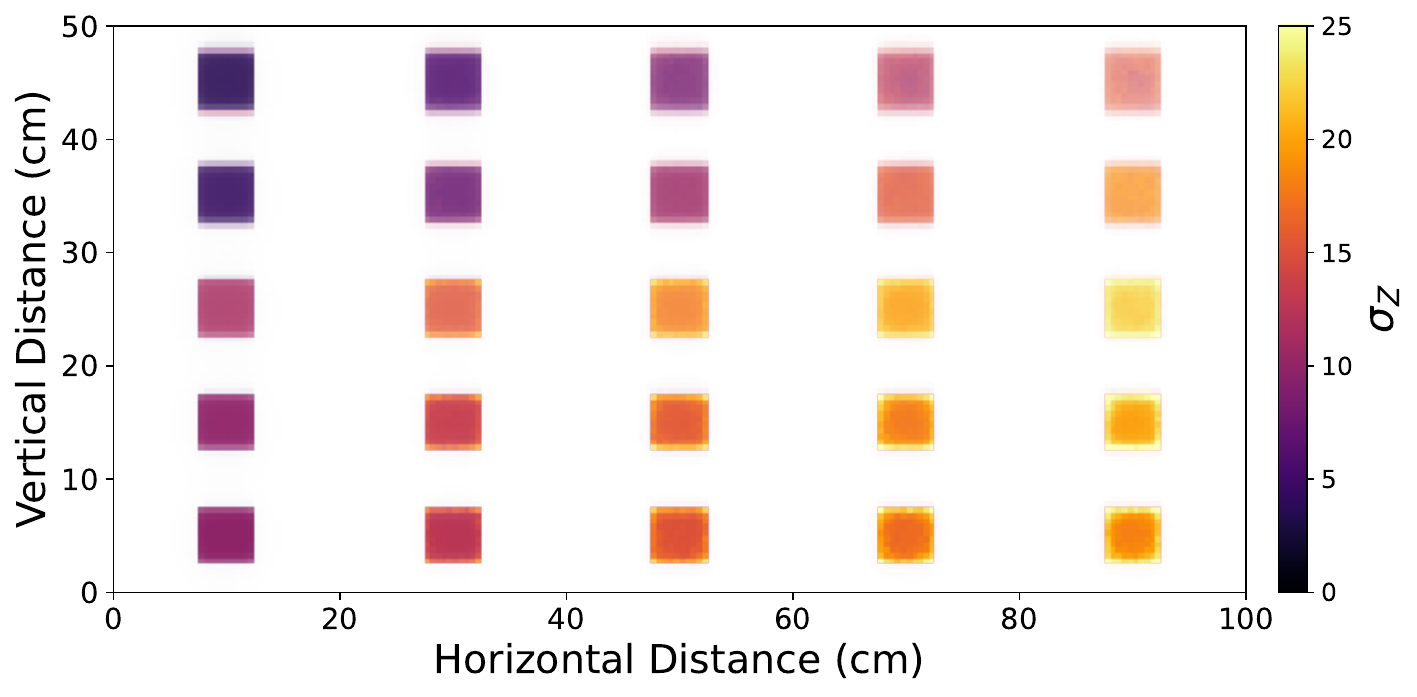}
        \caption{Atomic number reconstruction uncertainty of the noisy simulated shielded phantom after stripping the steel, computed as the pixel-wise sample standard deviation of reconstructed $Z$ output across 1000 runs.}
        \label{Z_shielded_peeled_unc}
    \end{subfigure}
    \caption{Atomic number reconstruction results using the simulated radiographic shielded phantom.}
\end{figure}

To explore the effects of shielding on the ability to reconstruct the atomic number of different materials, we simulate a shielded radiographic phantom. We place four vertical steel slabs with area densities ranging from $50 \g/\cm^2$ to $200 \g/\cm^2$ (thicknesses of $6.4 \cm$ to $25.4 \cm$). We then include five rows of materials by placing graphite $(Z=6)$, aluminum $(Z=13)$, tin $(Z=50)$, lead $(Z=82)$, and plutonium $(Z=94)$ behind the slabs of steel. We also include one column with no steel, corresponding to an unshielded measurement. This produces a $5 \times 5$ matrix geometry, where each row represents a different material, and each column represents a different thickness of steel shielding. The simulation geometry is described further in \ref{simulation_geom}. Fig.~\ref{Z_shielded_peeled_true} shows the ground truth $\{\lambda, Z\}$ map of the objects behind the steel shielding, and Fig.~\ref{simulation_H_shielded_noisy} shows the simulated image using the noise model described in Section~\ref{Noise_model}.

The two pass methodology of Section~\ref{identification_shielded} was then applied independently to 1000 noisy shielded phantom images. Figs.~\ref{Z_shielded_med} and \ref{Z_shielded_unc} show the first pass predicted atomic number and uncertainty, respectively. We observe that during the first pass, the presence of the steel shielding significantly suppresses the ability to identify materials by their $Z$. Next, using the output of the first pass to approximate $\lambda_\text{shield}$ and $Z_\text{shield}$, we perform a second pass, mathematically stripping the steel. Figs. \ref{Z_shielded_peeled_med} and \ref{Z_shielded_peeled_unc} show the predicted atomic number and uncertainty output of the second pass, and these results are quantified in table \ref{table_shielded}. Using this two pass approach, we are able to obtain atomic number estimates which are consistent with the ground truth $Z$ of the unshielded objects, despite the thick shielding present in the images. Even with $25.4\cm$ of steel, we are able to classify graphite as low-$Z$, and lead and plutonium as high-$Z$. This offers a potential avenue for dual energy cargo inspection systems to identify shielded high-$Z$ materials.

We observe relatively large uncertainties for the heavily shielded materials, sometimes as high as $\sigma_Z \approx 25$. This result stresses that in order to obtain precise confidence bounds for heavily shielded materials, it is necessary to use sufficiently long measurement times to obtain adequate statistics.

\begin{table}
\begin{centering}
\begin{tabular}{l c c c c c c}
\toprule
 & Ground Truth $Z_\eff$ & No shielding & $50\g/\cm^2$ steel & $100\g/\cm^2$ steel & $150\g/\cm^2$ steel & $200\g/\cm^2$ steel \\
\midrule
 Graphite & $6$ & $9 \pm 4$ & $7 \pm 6$ & $7 \pm 8$ & $6 \pm 12$ & $4 \pm 15$ \\
 Aluminum & $13$ & $15 \pm 4$ & $16 \pm 8$ & $14 \pm 11$ & $15 \pm 15$ & $12 \pm 18$ \\
 Tin & $50$ & $52 \pm 11$ & $52 \pm 15$ & $51 \pm 18$ & $46 \pm 20$ & $42 \pm 22$ \\
 Lead & $\{82, 87\}$ & \makecell{$\{76 \pm 10$, \\[-1.5ex]~ $84 \pm 11\}$} & \makecell{$\{72 \pm 13$, \\[-1.5ex]~ $88 \pm 17\}$} & \makecell{$\{71 \pm 16$, \\[-1.5ex]~ $90 \pm 20\}$} & \makecell{$\{67 \pm 18$, \\[-1.5ex]~ $91 \pm 23\}$} & \makecell{$\{63 \pm 20$, \\[-1.5ex]~ $91 \pm 26\}$} \\
 Plutonium & $\{74, 94\}$ & \makecell{$\{79 \pm 10$, \\[-1.5ex]~ $84 \pm 10\}$} & \makecell{$\{78 \pm 13$, \\[-1.5ex]~ $88 \pm 14\}$} & \makecell{$\{74 \pm 15$, \\[-1.5ex]~ $91 \pm 18\}$} & \makecell{$\{73 \pm 17$, \\[-1.5ex]~ $91 \pm 20\}$} & \makecell{$\{71 \pm 19$, \\[-1.5ex]~ $91 \pm 22\}$} \\
\bottomrule
\end{tabular}
\caption{Reconstructing the $Z$ of different objects placed behind a steel shield. Mathematically stripping off the steel shield allows for accurate identification of the shielded object. Cells with two entries indicate the presence of a $Z$ degeneracy in which multiple solutions were found.}
\label{table_shielded}
\end{centering}
\end{table}

\section{Conclusion}
\label{Conclusion}

This work introduces a flexible algorithm for predicting the area density and atomic number of dual energy radiographic images. We introduce a semiempirical transparency model and choose the optimal $Z_\eff$ by minimizing a chi-squared loss function. We identify two principle advantages of our method: first, the calibration step used in this study requires only three scans, enabling simpler and more frequent system recalibrations. Second, the $Z_\eff$ output of this approach is unbinned, removing bias due to material binning and reducing interpolation error.

In this study, we provide a proof of concept in Geant4 by simulating two radiographic phantom images. We are able to calculate atomic number estimates which are consistent with ground truth, even using noisy input images and shielded materials. The authors acknowledge that this result is not a final proof that this method will work in a real system. We outline a clear procedure for porting this approach to commercial systems, and recommend that future work should focus on experimentally testing this method in deployed radiographic imaging scenarios.

\section{Acknowledgements}

This work was supported by the Department of Energy Computational Science Graduate Fellowship (DOE CSGF) under grant DE-SC0020347. The authors would like to acknowledge Cristian Dinca at Rapiscan Systems for his useful suggestions and feedback. The authors declare no conflict of interest.
\newpage

\bibliography{References.bib}

\newpage

\appendix

\section{}
\label{appendix}

\subsection{Simulation of the beam spectra and detector response}
\label{simulation}

\begin{figure}
\begin{centering}
\includegraphics[width=0.49\textwidth]{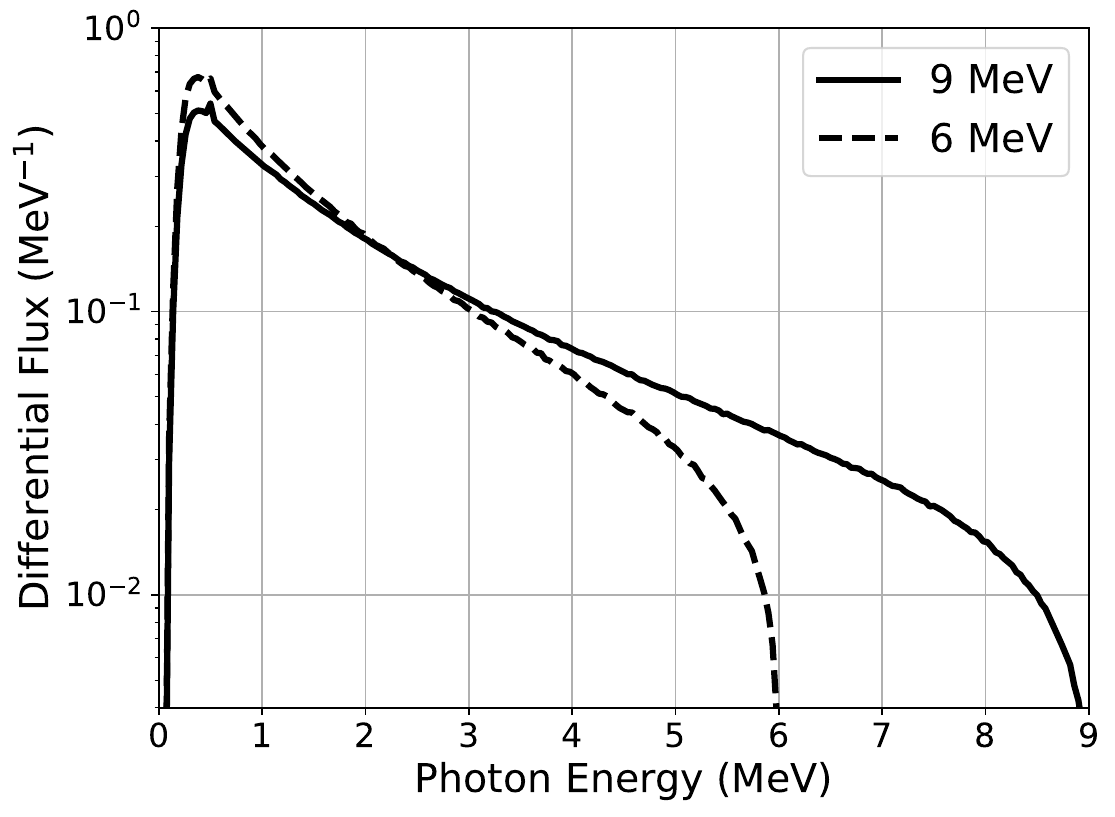}
\caption{Simulated $9$ and $6$ MeV endpoint energy bremsstrahlung beam spectra ($\phi_H$ and $\phi_L$, respectively).}
\label{spectra}
\end{centering}
\end{figure}

The incident photon beam spectra and detector response function were simulated using Geant4 using the QGSP BIC physics list with a production cut of 80 keV~\cite{Geant4}. To simulate the bremsstrahlung beam spectra, $9$ and $6$ MeV electrons were directed at a 0.1cm tungsten radiator backed by 1cm of copper. An additional 1cm of steel was included to filter the low energy X-rays. The resulting bremsstrahlung photons were subsequently measured by a tally surface and binned according to their beam angle. $\phi_{\{H, L\}}$ were calculated from the smallest angular bin, subtending a half angle of ${\approx} 0.3^{\circ}$. The resulting dual energy beam spectra are shown in Fig.~\ref{spectra}. This geometry was designed to be simple, generalizable, and representative of typical cargo scanning applications. While the beam filtration used by real scanning systems is much more sophisticated, this configuration is intended for algorithmic benchmark purposes.

To simulate the detector response, photons with energy uniformly distributed between $0$ and $9$ MeV were directed along the long axis of a $1.5 \times 1.0 \times 3.0 \cm$ cadmium tungstate (CdWO$_4$) crystal. The incident energy and total deposited energy of each photon was recorded and binned to produce the detector response function.

\subsection{Calculation of the calibration parameters}
\label{calculating_abc}

In radiographic systems, the presence of scattering causes the detected transparency to diverge from an idealized free streaming photon model. Past work showed that these bulk secondary effects can be corrected for using the semiempirical log-transparency model, given by Eq.~\ref{alpha}~\cite{Lalor2023}:

\begin{align}
\begin{split}
&\tilde \alpha_H (\lambda, Z) = -\log \frac{\int_0^{\infty} D(E) \phi_H(E) e^{-\tilde \mu_H (E, Z) \lambda} dE}{\int_0^{\infty}D(E) \phi_H(E) dE} \\
&\tilde \alpha_L (\lambda, Z) = -\log \frac{\int_0^{\infty} D(E) \phi_L(E) e^{-\tilde \mu_L (E, Z) \lambda} dE}{\int_0^{\infty}D(E) \phi_L(E) dE}
\end{split}
\label{alpha}
\end{align}

\begin{figure}
    \centering
    \includegraphics[width=0.47\textwidth]{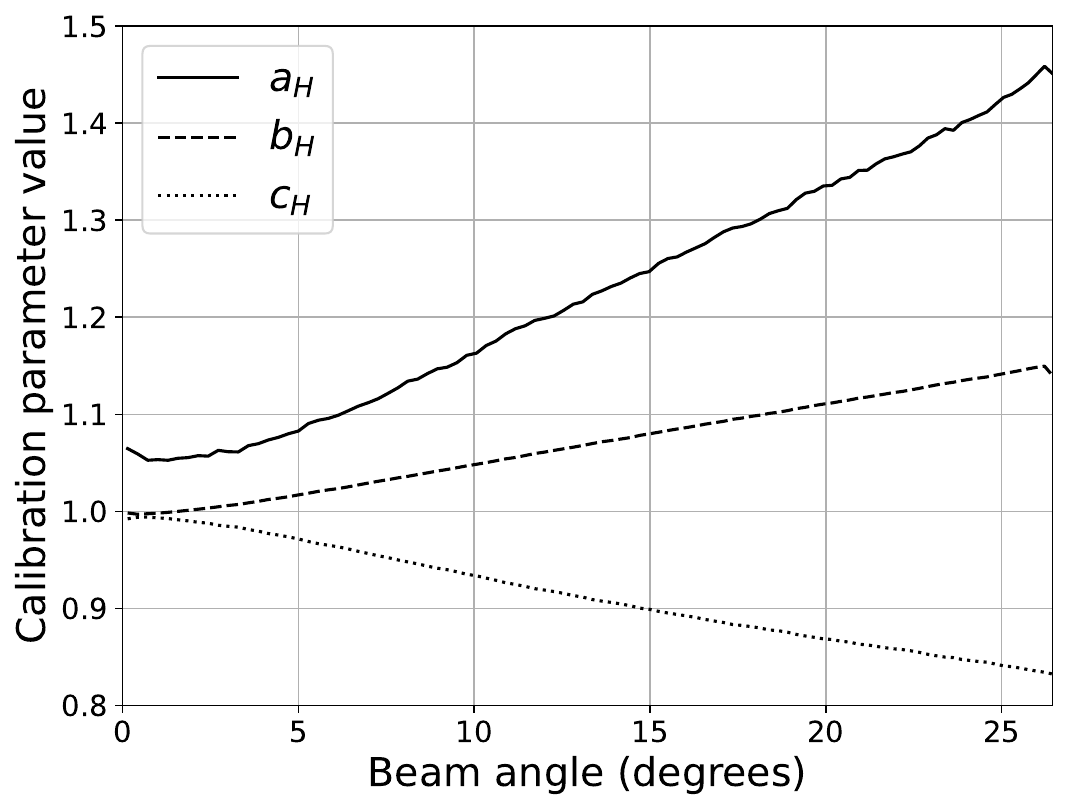}
    \hfill
    \includegraphics[width=0.47\textwidth]{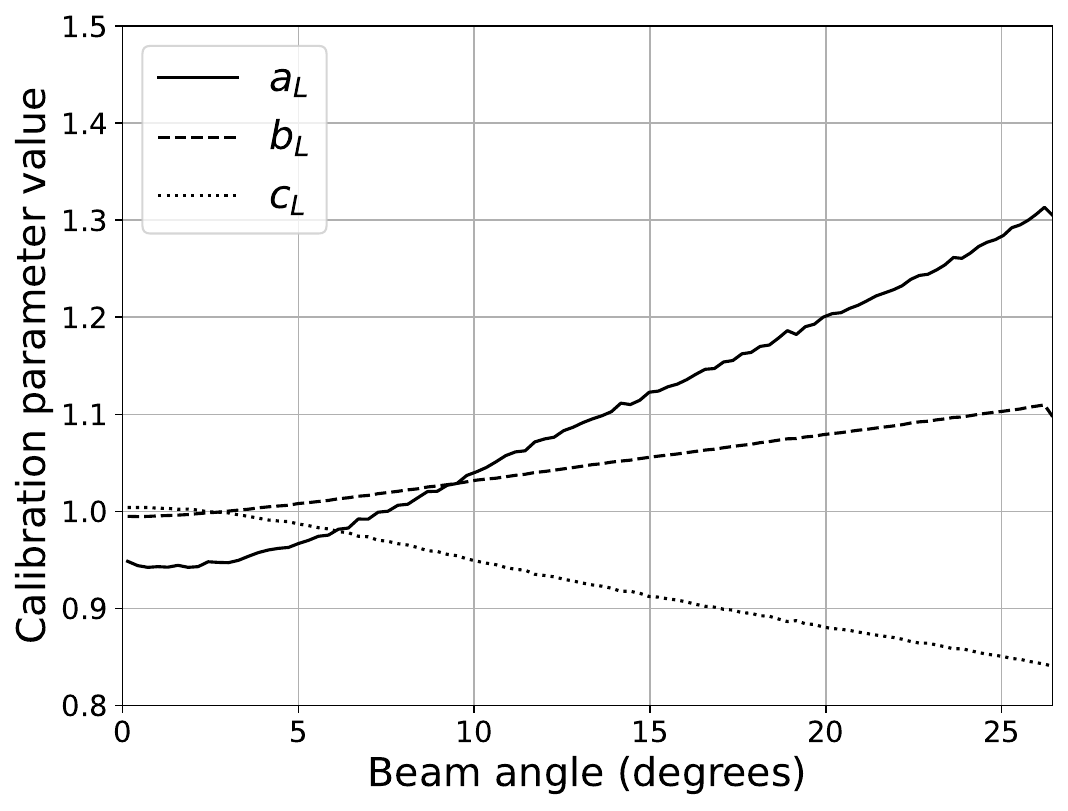}
    \caption{Values of the calibration parameters $a$, $b$, and $c$ as a function of the bremsstrahlung beam angle, shown for the $9$ MeV beam (left) and $6$ MeV beam (right). The trend in $a$, $b$, and $c$ can be explained by the property that at larger detector angles, the bremsstrahlung beam shifts towards lower energies.}
    \label{calibration}
\end{figure}

To calibrate the model, high resolution transparency simulations of graphite $(Z=6)$, iron $(Z=26)$, and lead $(Z=82)$ were performed at an area density of $\lambda = 100 \g/\cm^2$. Then, $a$, $b$, and $c$ were determined by minimizing the squared error between the transparency model and transparency simulations:

\begin{equation}
\begin{split}
a_H, b_H, c_H &= \argmin_{a, b, c} \sum_i \left(\tilde \alpha_H(\lambda_i, Z_i) - \alpha_{H,i} \right)^2 \\
a_L, b_L, c_L &= \argmin_{a, b, c} \sum_i \left(\tilde \alpha_L(\lambda_i, Z_i) - \alpha_{L,i} \right)^2
\end{split}
\label{calc_abc}
\end{equation}

where $\{\alpha_{H,i}, ~\alpha_{L,i}\}$ are the measured log-transparencies in the presence of calibration material $\{\lambda_i, Z_i\}$. In Eq.~\ref{calc_abc}, the transparency models $\tilde \alpha_H(\lambda, Z)$ and $\tilde \alpha_L(\lambda, Z)$ (Eq.~\ref{alpha}) depend implicitly on $a$, $b$, and $c$ through the semiempirical mass attenuation coefficient (Eq.~\ref{semiempirical_mass_atten}). The values for the calibration parameters used in this study are shown in Fig.~\ref{calibration}. Each detector is calibrated separately, and thus $a$, $b$, and $c$ depend on the detector index. At least three calibration measurements are necessary in order for Eq.~\ref{calc_abc} to yield a unique solution. This analysis found that including more than three calibration measurements yielded diminishing improvements.

\subsection{Calculating ground truth}
\label{calc_ground_truth_appendix}

To approximate the ground truth $Z_\eff$ of each pixel, we wish to select the homogeneous material which best represents the array of materials along the beam path. To do this, we first adapt Eq.~\ref{alpha} for the heterogeneous case: 

\begin{equation}
\begin{split}
&\tilde \alpha_H (\vec \lambda, \vec Z) = -\log \frac{\int_0^{\infty} D(E) \phi_H(E) e^{-\sum_j \tilde \mu_H (E, Z_j) \lambda_j} dE}{\int_0^{\infty}D(E) \phi_H(E) dE} \\
&\tilde \alpha_L (\vec \lambda, \vec Z) = -\log \frac{\int_0^{\infty} D(E) \phi_L(E) e^{-\sum_j \tilde \mu_L (E, Z_j) \lambda_j} dE}{\int_0^{\infty}D(E) \phi_L(E) dE}
\end{split}
\label{alpha_heterogeneous}
\end{equation}

where the photon beam passes through an array of materials $\vec Z$ with corresponding area densities $\vec \lambda$. For this work, we calculate the effective atomic number and area density of a known $\{\vec \lambda, \vec Z\}$ as the solution to Eq.~\ref{ground_truth}:

\begin{equation}
\lambda_\eff, ~Z_\eff = \solve\limits_{\lambda, Z}
\begin{cases}
	\tilde \alpha_H(\lambda, Z) = \tilde \alpha_H(\vec \lambda, \vec Z) \\
	\tilde \alpha_L(\lambda, Z) = \tilde \alpha_L(\vec \lambda, \vec Z) \\
\end{cases}
\label{ground_truth}
\end{equation}

Unfortunately, the solution to Eq.~\ref{ground_truth} is not guaranteed to be unique. This property is due to a fundamental shortcoming of dual energy radiographic systems in which two different materials can sometimes produce identical transparency measurements~\cite{Lalor2024}. As a result, certain measurements yield a solution degeneracy which cannot be resolved through better algorithms or improved statistics. For this work, in the case of non-unique solutions to Eq.~\ref{ground_truth}, we record both atomic number solutions.

\subsection{Performing the chi-squared optimization}
\label{optimization}

We compute the $\{\lambda, Z\}$ of every pixel in a radiographic image by minimizing Eq.~\ref{chi2}. While Eq.~\ref{chi2} is not guaranteed to be convex for arbitrary beam spectra and detector response, we numerically verify that it is strictly convex in the $\boldsymbol{\lambda}$ dimension for the parameters used in this study, and we expect this property to be true for any realistic application. As a result, for a fixed $Z$, we can quickly minimize Eq.~\ref{chi2} using only a few iterations of Newton's method:

\begin{equation}
\boldsymbol{\lambda}^{(n+1)} = \boldsymbol{\lambda}^{(n)} - \left[\frac{\partial^2 \chi^2 (\boldsymbol{\lambda}^{(n)}, Z)}{\partial \boldsymbol{\lambda}^2}\right]^{-1} \frac{\partial \chi^2 (\boldsymbol{\lambda}^{(n)}, Z)}{\partial \boldsymbol{\lambda}}
\label{newton}
\end{equation}

Expressions for the gradient and Hessian matrix are given in \ref{calculus}. The Hessian matrix is diagonal, so the inversion and subsequent matrix multiplication can be performed in $\mathcal{O}(|C|)$ time, where $|C|$ is the number of pixels in the segment. We repeat this procedure exhaustively for $1 \leq Z \leq 100$ to find the global minimum of Eq.~\ref{chi2}. We can speed up the calculation by noticing that the converged value of $\boldsymbol{\lambda}$ in Eq.~\ref{newton} for a given $Z$ is similar to the result for neighboring $Z$ values. Thus, we can use the converged $\boldsymbol{\lambda}$ for $Z=1$ as the initial guess when calculating $\boldsymbol{\lambda}$ for $Z = 2$, and so on. Under this procedure, after solving the $Z=1$ case, we can solve $Z=2\dots100$ each using a single Newton step. We perform a runtime analysis of the method in \ref{runtime analysis}, finding that minimizing Eq.~\ref{chi2} for every segment in a typical radiograph containing two million pixels took less than $30$ seconds using a single core of an Intel Core i7.

Sometimes, Eq.~\ref{chi2} admits two local minima. This stems from a fundamental limitation of dual energy radiographic systems in which two different materials can  produce identical transparency measurements~\cite{Lalor2024}. For this work, if multiple local optima are found, we record both atomic number solutions.

\subsection{Computing the gradient and Hessian matrix}
\label{calculus}

In order to calculate the gradient and the Hessian matrix (for computing Eq.~\ref{newton}), we first differentiate Eq.~\ref{chi2} with respect to $\boldsymbol{\lambda}$. To simplify notation, we will drop index labels and evaluate the derivatives elementwise. Using the chain rule:

\begin{equation}
\frac{\partial \chi^2(\lambda, Z)}{\partial \lambda} = \frac{2}{\sigma_{\alpha_H}^2} \left( \tilde \alpha_H(\lambda, Z) - \alpha_H \right) \frac{\partial \tilde \alpha_H(\lambda, Z)}{\partial \lambda} + \frac{2}{\sigma_{\alpha_L}^2} \left( \tilde \alpha_L(\lambda, Z) - \alpha_L \right) \frac{\partial \tilde \alpha_L(\lambda, Z)}{\partial \lambda}
\label{chi2_d1}
\end{equation}

and the second derivative:

\begin{equation}
\begin{split}
\frac{\partial^2 \chi^2(\lambda, Z)}{\partial \lambda^2} &= \frac{2}{\sigma_{\alpha_H}^2} \left( \tilde \alpha_H(\lambda, Z) - \alpha_H \right) \frac{\partial^2 \tilde \alpha_H(\lambda, Z)}{\partial \lambda^2} + \frac{2}{\sigma_{\alpha_H}^2} \left( \frac{\partial \tilde \alpha_H(\lambda, Z)}{\partial \lambda} \right)^2 \\
&~+ \frac{2}{\sigma_{\alpha_L}^2} \left( \tilde \alpha_L(\lambda, Z) - \alpha_L \right) \frac{\partial^2 \tilde \alpha_L(\lambda, Z)}{\partial \lambda^2} + \frac{2}{\sigma_{\alpha_L}^2} \left( \frac{\partial \tilde \alpha_L(\lambda, Z)}{\partial \lambda} \right)^2
\end{split}
\label{chi2_d2}
\end{equation}

We note that the cross terms of the second derivative are zero:

\begin{equation}
\frac{\partial^2 \chi^2(\boldsymbol{\lambda}, Z)}{\partial \lambda_i \partial \lambda_j} = 0~~,~~i \neq j
\label{off-diagonal}
\end{equation}

and thus the Hessian is diagonal. To calculate $\frac{\partial {\tilde \alpha}(\lambda, Z)}{\partial \lambda}$ and $\frac{\partial^2 {\tilde \alpha}(\lambda, Z)}{\partial \lambda^2}$, we differentiate Eq.~\ref{alpha} with respect to $\boldsymbol{\lambda}$, yielding:

\begin{equation}
\begin{split}
\frac{\partial \tilde \alpha_H(\lambda, Z)}{\partial \lambda} &= \frac{\int_0^{\infty} D(E) \phi_H(E) \tilde \mu_H (E, Z) e^{-\tilde \mu_H (E, Z) \lambda} dE}{\int_0^{\infty} D(E) \phi_H(E) e^{-\tilde \mu_H (E, Z) \lambda} dE} \\
\frac{\partial \tilde \alpha_L(\lambda, Z)}{\partial \lambda} &= \frac{\int_0^{\infty} D(E) \phi_L(E) \tilde \mu_L (E, Z) e^{-\tilde \mu_L (E, Z) \lambda} dE}{\int_0^{\infty} D(E) \phi_L(E) e^{-\tilde \mu_L (E, Z) \lambda} dE}
\end{split}
\label{alpha_d1}
\end{equation}

\begin{equation}
\begin{split}
\frac{\partial^2 \tilde \alpha_H(\lambda, Z)}{\partial \lambda^2} &= \left(\frac{\int_0^{\infty} D(E) \phi_H(E) \tilde \mu_H (E, Z) e^{-\tilde \mu_H (E, Z) \lambda} dE}{\int_0^{\infty} D(E) \phi_H(E) e^{-\tilde \mu_H (E, Z) \lambda} dE}\right)^2 - \frac{\int_0^{\infty} D(E) \phi_H(E) \tilde \mu_H (E, Z)^2 e^{-\tilde \mu_H (E, Z) \lambda} dE}{\int_0^{\infty} D(E) \phi_H(E) e^{-\tilde \mu_H (E, Z) \lambda} dE} \\
\frac{\partial^2 \tilde \alpha_L(\lambda, Z)}{\partial \lambda^2} &= \left(\frac{\int_0^{\infty} D(E) \phi_L(E) \tilde \mu_L (E, Z) e^{-\tilde \mu_L (E, Z) \lambda} dE}{\int_0^{\infty} D(E) \phi_L(E) e^{-\tilde \mu_L (E, Z) \lambda} dE}\right)^2 - \frac{\int_0^{\infty} D(E) \phi_L(E) \tilde \mu_L (E, Z)^2 e^{-\tilde \mu_L (E, Z) \lambda} dE}{\int_0^{\infty} D(E) \phi_L(E) e^{-\tilde \mu_L (E, Z) \lambda} dE}
\end{split}
\label{alpha_d2}
\end{equation}

Prior to any minimization, lookup tables of $\tilde \alpha_H(\lambda, Z)$, $\tilde \alpha_L(\lambda, Z)$, $\frac{\partial \tilde \alpha_H(\lambda, Z)}{\partial \lambda}$, $\frac{\partial \tilde \alpha_L(\lambda, Z)}{\partial \lambda}$, $\frac{\partial^2 \tilde \alpha_H(\lambda, Z)}{\partial \lambda^2}$, and $\frac{\partial^2 \tilde \alpha_L(\lambda, Z)}{\partial \lambda^2}$ (Eqs. \ref{alpha}, \ref{alpha_d1}, and \ref{alpha_d2}) were made for an array of $\lambda$ and $Z$ values. This was done to make the calculations of Eqs. \ref{chi2}, \ref{chi2_d1}, and \ref{chi2_d2} during minimization significantly cheaper since, instead of numerical integrals, only table lookups are required. Since each detector was calibrated independently, different lookup tables were made for different detectors.

\subsection{Runtime analysis}
\label{runtime analysis}

\begin{figure}
\begin{centering}
\includegraphics[width=0.49\textwidth]{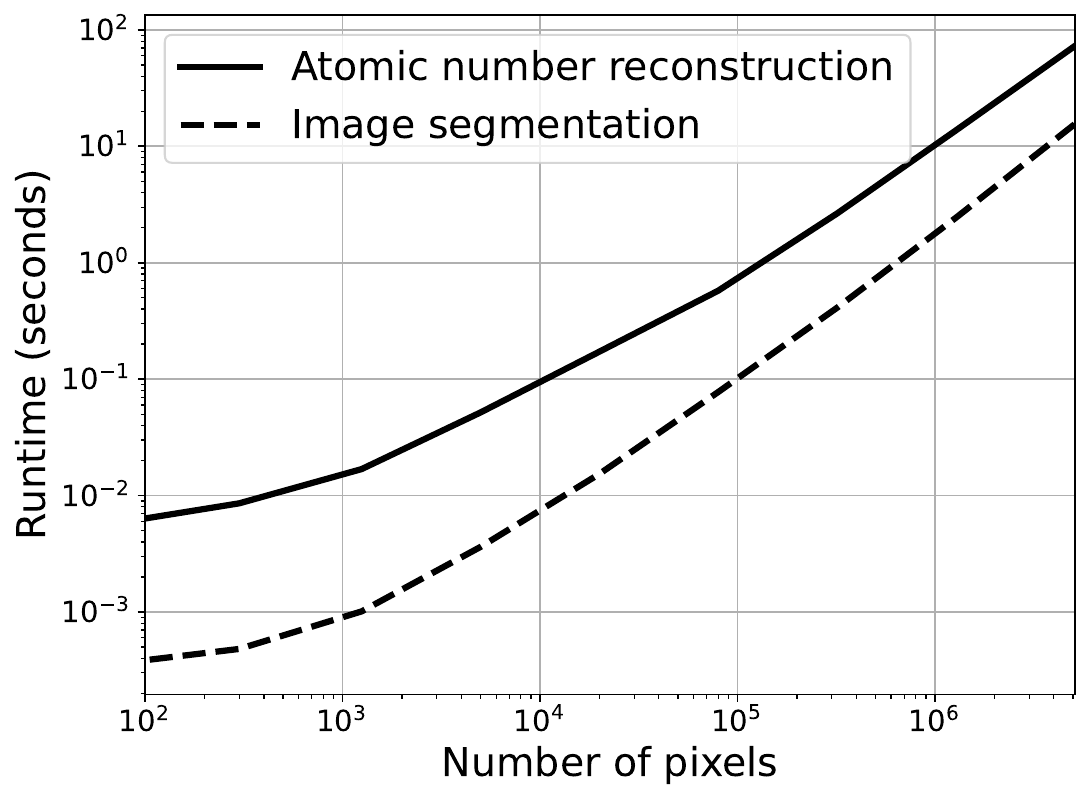}
\caption{Runtime of the atomic number reconstruction algorithm and image segmentation routine as a function of the size of the input image. We see approximately linear scaling with image size and verify that the algorithm is fast enough for real-time security applications.}
\label{runtime}
\end{centering}
\end{figure}

The feasibility of any atomic number reconstruction routine is dependent on the ability to be run in real-time for practical cargo security applications. To explore this effect, we upsampled the radiographic cargo phantom and recorded the runtime of both the atomic number reconstruction algorithm and the image segmentation routine as a function of image size. The results are shown in Fig.~\ref{runtime}. We verify that both the image segmentation step and the atomic number reconstruction step scale approximately linearly with the number of pixels. Furthermore, we note that the total runtime is independent of the number of pixel segments. The analysis in this section was performed using a single core of an Intel Core i7.

\subsection{Simulating radiographic phantom images}
\label{simulation_geom}

\begin{figure}
    \centering
    \begin{subfigure}[t]{0.49\textwidth}
        \centering
        \includegraphics[width=\textwidth]{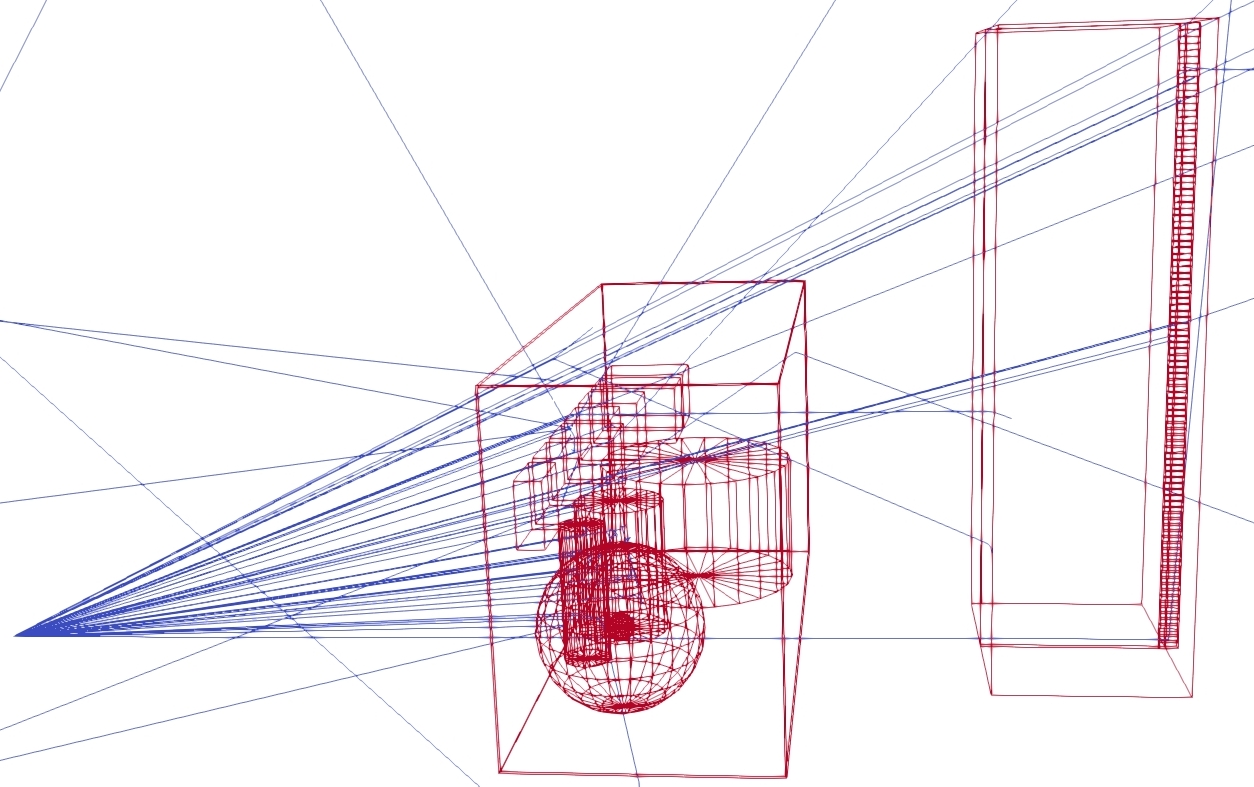}
        \caption{Radiographic cargo phantom simulation geometry}
        \label{cargo_geom}
    \end{subfigure}
    \begin{subfigure}[t]{0.49\textwidth}
        \centering
        \includegraphics[width=\textwidth]{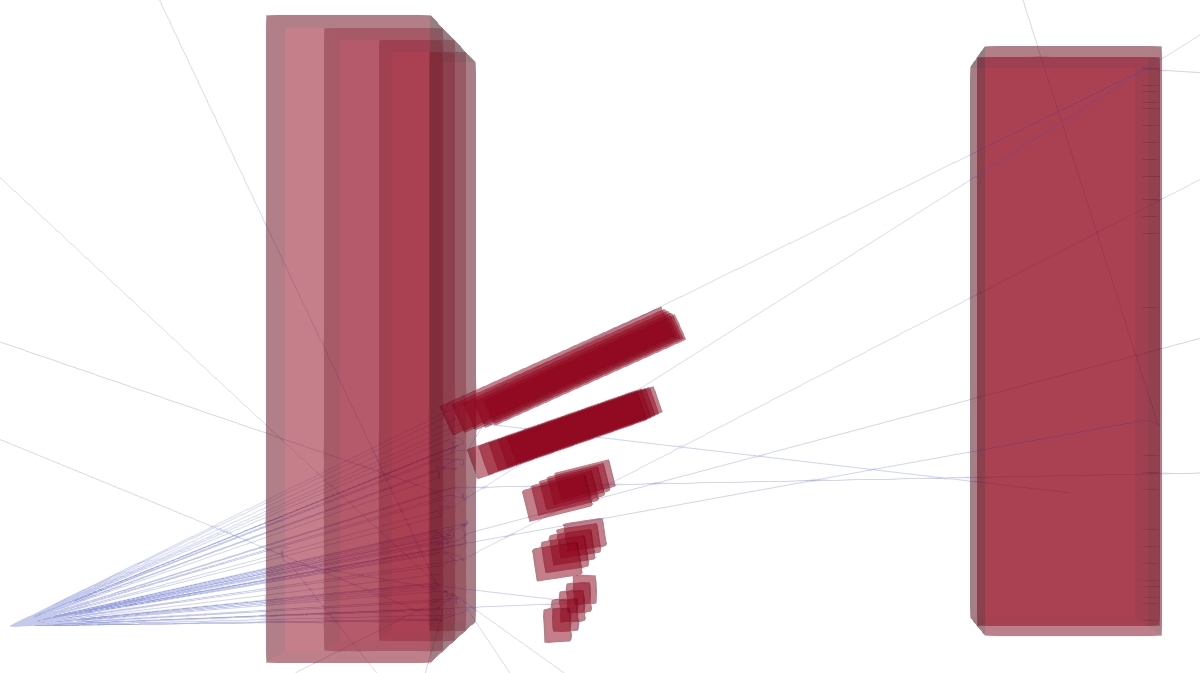}
        \caption{Radiographic shielded phantom simulation geometry}
        \label{shielded_geom}
    \end{subfigure}
    \caption{Simulation geometry of the different radiographic phantoms considered in this study}
    \label{geom}
\end{figure}

This work simulated two radiographic phantoms to test the performance of the atomic number reconstruction routine. The radiographic cargo phantom consisted of a steel container with a thickness of $0.2 \cm$. Inside the container were seven boxes composed of graphite ($Z=6,~ \lambda=30 \g/\cm^2$), aluminum ($Z=13,~ \lambda=40 \g/\cm^2$), iron ($Z=26,~ \lambda=79 \g/\cm^2$), silver ($Z=47,~ \lambda=79 \g/\cm^2$), gadolinium ($Z=64,~ \lambda = 79 \g/\cm^2$), lead ($Z=82,~ \lambda= 79 \g/\cm^2$), and uranium ($Z=92,~ \lambda=76 \g/\cm^2$). The boxes are all $10\cm \times 10\cm$ and placed $30\cm$ above the floor of the container. Below the boxes are cylinders of water (H$_2$O, $r = 17.5 \cm$), silver chloride (AgCl,  $r=7.5 \cm$), and uranium oxide (UO$_2$, $r=3.75 \cm$), all with a height of $22.5\cm$. Next to the cylinders is a plutonium pit ($Z=94,~ r=2\cm$), surrounded by a shell of polyethylene shielding (CH$_2$, $r=12.5 \cm$). The simulation geometry is shown in Fig.~\ref{cargo_geom}.

To explore the effects of shielding, we also simulated a radiographic shielded phantom. We created a $5 \times 5$ grid, placing five graphite boxes ($Z=6,~ \lambda = 76 \g/\cm^2$) in the top row, five aluminum boxes ($Z=13,~ \lambda = 76 \g/\cm^2$) in the second row, five tin boxes ($Z=50,~ \lambda = 73 \g/\cm^2$) in the third row, five lead boxes ($Z=82,~ \lambda = 79 \g/\cm^2$) in the fourth row, and five plutonium boxes ($Z=94,~ \lambda = 79 \g/\cm^2$) in the bottom row. The boxes were all $5\cm \times 5\cm$ and angled towards the beam source to minimize edge effects. In the first column, we include no shielding other than air. In the second through fifth columns, we add steel slabs with area densities of $50 \g/\cm^2$, $100 \g/\cm^2$, $150 \g/\cm^2$, and $200 \g/\cm^2$ (corresponding to thicknesses from $6.4 \cm$ to $25.4 \cm$). The simulation geometry is shown in Fig.~\ref{shielded_geom}. This geometry was designed to benchmark the guidelines set forth by the Domestic Nuclear Detection Office (DNDO), requiring detection of $100 \cm^3$ of high-$Z$ material $(Z \geq 72)$ behind $25.4 \cm$ of steel~\cite{Bentley2012}.

To perform radiography simulations, photons were directed in a fan beam towards one meter tall stack of CdWO$_4$ detectors with a $30$cm lead collimator to filter scattered radiation. The detector stack was placed two meters from the photon source, and the target was placed midway between the source and the detector. The photon beam was divided into $100$ angular bins to capture the angular dependence of bremsstrahlung photon energy. For each beam angular bin, photons were sampled from the corresponding energy distribution calculated in \ref{simulation}. Each simulation, the target was shifted by $0.5\cm$, and this was repeated across the entire length of the target in order to construct the full 2D radiograph.

\end{sloppypar}
\end{document}